\documentclass{emulateapj}
\shorttitle{Star cluster disruption in M82}
\shortauthors{Shuo Li et al.}
\slugcomment{Submitted for publication in The Astrophysical Journal Supplements}

\begin{document}

\title{Star cluster disruption in the starburst galaxy Messier 82}

\author{Shuo Li\altaffilmark{1,2}, Richard de Grijs\altaffilmark{2,1},
  Peter Anders\altaffilmark{3}, and Chengyuan Li\altaffilmark{1,2}}
\altaffiltext{1}{Department of Astronomy, Peking University, Yi He
  Yuan Lu 5, Hai Dian District, Beijing 100871, China} 
\altaffiltext{2}{Kavli Institute for Astronomy \& Astrophysics, Peking
  University, Yi He Yuan Lu 5, Hai Dian District, Beijing 100871,
  China; grijs@pku.edu.cn}
\altaffiltext{3} {Key Laboratory for Optical Astronomy, National
  Astronomical Observatories, Chinese Academy of Sciences, 20A Datun
  Road, Chaoyang District, Beijing 100012, China}

\begin{abstract}
Using high-resolution, multiple-passband {\sl Hubble Space Telescope}
images spanning the entire optical/near-infrared wavelength range, we
obtained a statistically complete sample, $U$-band selected sample of
846 extended star clusters across the disk of the nearby starburst
galaxy M82. Based on careful analysis of their spectral energy
distributions, we determined their galaxy-wide age and mass
distributions. The M82 clusters exhibit three clear peaks in their age
distribution, thus defining a relatively young, $\log( t \mbox{
  yr}^{-1})\leq7.5$, an intermediate-age, $\log( t \mbox{
  yr}^{-1})\in[7.5, 8.5]$, and an old sample, $\log( t \mbox{
  yr}^{-1}) \geq8.5$. Comparison of the completeness-corrected mass
distributions offers a firm handle on the galaxy's star cluster
disruption history. The most massive star clusters in the young and
old samples are (almost) all concentrated in the most densely
populated central region, while the intermediate-age sample's most
massive clusters are more spatially dispersed, which may reflect the
distribution of the highest-density gas throughout the galaxy's
evolutionary history, combined with the solid-body nature of the
galaxy's central region.
\end{abstract}

\section{Introduction}

Most current star-formation scenarios are based on the notion that
70\%--90\% of stars form in embedded clusters (e.g., Lada \& Lada
2003). In their comprehensive review, Portegies Zwart et al. (2010)
point out that this is, in fact, supported by the global clustering of
spectral O-type stars (Parker \& Goodwin 2007), of which $\sim$70\%
reside in young clusters or associations (Gies 1987) and $\sim$50\% of
the remaining field population are directly identified as runaways (de
Wit et al. 2005). Moreover, even some of the merely $\sim$4\% of
O-type stars that may not have formed within a clustered environment
might indeed also be runaway stars (Gvaramadze \& Bomans 2008;
Schilbach \& R\"oser 2008; Pflamm-Altenburg \& Kroupa 2010). A
comparison of the observed formation rate of stars in embedded
clusters ($\sim 3 \times 10^3 M_{\odot}$ Myr$^{-1}$ kpc$^{-2}$; Lada
\& Lada 2003) with stars in the field ($\sim 3$--$7 \times 10^3
M_{\odot}$ Myr$^{-1}$ kpc$^{-2}$; Miller \& Scalo 1979) also strongly
supports the notion that clusters embody the fundamental mode of star
formation.\footnote{Some level of disagreement persists as to whether
  stars physically form in star clusters or in more diffuse structures
  (`fractal distributions,' following the turbulent structure of the
  interstellar medium; e.g., Gieles et al. 2008; Bastian et al. 2009,
  2011; Bressert et al. 2010; Kruijssen 2012; Longmore et al. 2014)
  that collapse into cluster-like configurations within a few Myr.}

\subsection{Young Massive Star Clusters}

Young massive star clusters (YMCs) are compact regions of very active
star formation. Based on analyses of their derived masses, radii, and
ages, it is thought that YMCs, particularly the most massive and
brightest YMCs, could over a Hubble time probably evolve into a
population of globular clusters (GCs; e.g., de Grijs \& Parmentier
2007; Portegies Zwart et al. 2010, and references therein). 

But can this evolution happen? Meurer et al. (1995) and de Grijs \&
Parmentier (2007, and references therein) suggested that any such
evolution depends sensitively on the stellar mass distribution,
especially on the content of low-mass stars in YMCs. If the stellar
initial mass function (IMF) in the clusters is biased toward high-mass
stars (a `top-heavy' IMF), it is unlikely that the clusters can
survive the disruptive nature of mass loss due to stellar evolution
and dynamical processes (Chernoff \& Shapiro 1987; Chernoff \&
Weinberg 1990; Goodwin 1997; Takahashi \& Portegies Zwart 2000; Smith
\& Gallagher 2001; Mengel et al. 2002; Kouwenhoven et al. 2014, based
on numerical simulations). 

YMCs are important objects to study because of the insights they may
provide into the formation and destruction of (proto-)GCs (on which we
focus in this paper), the star-formation processes in extreme
environments, and the triggering and feeding of supergalactic winds
(Whitmore et al. 1999; Larsen 2002; Scheepmaker et al. 2007; Hwang \&
Lee 2008, 2010; Cantiello et al. 2009; Chandar et al. 2010; Pellerin
et al. 2010; Whitmore et al. 2010; Larsen et al. 2011). In addition,
their ages and masses can be determined individually through either
spectroscopic or multi-passband imaging observations. Hence, they can
be used as powerful tracers of the starburst history across a given
galaxy. Fortunately, the high spatial resolution of {\sl Hubble Space
  Telescope} ({\sl HST}) observations makes the detection of compact
star clusters in starburst regions of several, relatively nearby
galaxies feasible.

\subsection{Messier 82}

Recent observations of M82 with the {\sl HST}/Advanced Camera for
Surveys (ACS), covering the entire optical extent of the galaxy, offer
an excellent opportunity to investigate the role of disruption
processes on the cluster mass function (CMF). The shape of the CMF as
a function of cluster age is a key diagnostic tool in our quest to
understand the evolution of a cluster population in a given galaxy. As
one of the nearest galaxies containing a large population of YMCs, the
luminosity and size-distribution functions can be studied better in
this galaxy than in almost any other galaxy of similar type.

M82 exhibits a biconical, extended filamentary structure (Ohyama et
al. 2002). It is the prototype nearby starburst galaxy (e.g., Barker
et al. 2008, and references therein) and contains a complex system of
star clusters. The energy from these clusters drives the famous
H$\alpha$- and X-ray-bright, kiloparsec-scale superwind (Shopbell \&
Bland-Hawthorn 1998; Stevens et al. 2003; Strickland \& Heckman
2007). The M82 system has a very complex structure, which varies on
both large and small scales, and its dynamics are also complicated by
the inflows and outflows caused by the galaxy's bar and its ongoing
starburst.

M82 is a member of the Messier 81 (M81) group. Within the last
200--500 Myr, at least one tidal encounter with M81 occurred. This
interaction is thought to be responsible for the triggering of the
starburst activity in M82. The strong and varying gravitational
effects associated with the most recent M81/M82 flyby have caused the
star-formation activity in M82 to increase by an order of magnitude
compared with other, `normal' galaxies. A substantial amount of gas
was funneled into the galaxy's core, which resulted in a concentrated
starburst, together with a corresponding marked peak in the cluster
age distribution. de Grijs et al. (2001) suggested that this starburst
may have continued for up to $\sim$50 Myr at a rate of $\sim$10
$M_{\odot}$ yr$^{-1}$. The core clusters, both YMCs and their less
massive counterparts, may have formed during the last of two
subsequent starbursts ($\sim$4--6 Myr ago).

In the core of M82, the active starburst region spans a diameter of
500 pc. At optical wavelengths, there are four high-surface-brightness
regions or clumps (designated A, C, D, and E by O'Connell \& Mangano
1978), which correspond to known sources at X-ray, infrared (IR), and
radio wavelengths. As a result, from our vantage point they are the
least obscured starburst complexes. Region A has been studied by many
researchers. O'Connell \& Mangano (1978) and O'Connell et al. (1995)
suggested that it may be composed of a remarkable complex of YMCs with
very high continuum and emission-line surface brightnesses. Shopbell
\& Bland-Hawthorn (1998) indicated that the well-known, large-scale
bipolar outflow or superwind appears to be centered on regions A and
C. Using {\sl HST} archival data, Melo et al. (2005) found a total of
197 optically visible clusters in the starburst core of M82, with 86
of these residing in region A. Smith et al. (2006) and Westmoquette et
al. (2007) used {\sl HST}/Space Telescope Imaging Spectrograph
observations to explore the environment (i.e., the state of the
interstellar medium) of the starburst clusters in M82 A and confirm
their formation during the most recent starburst event.

Numerous authors have also investigated the properties of the star
clusters located outside the central regions. de Grijs et al. (2001,
2003a, 2005) derived photometric ages for the extended, `fossil'
starburst region B located 0.5--1 kpc north-east of the nucleus,
showing that a peak of the star-formation activity occurred at $\log(t
\mbox{ yr}^{-1}) = 9.0 \pm 0.4$. After a correction due to having
previously underestimated the detection limit for well-resolved
clusters, Smith et al. (2007) presented new {\sl HST}/ACS $UBVI$
photometry for 35 $U$-band-selected massive star clusters in the
post-starburst region B of M82. They found that the peak epoch of
cluster formation for this sample occurred about 150 Myr ago and star
formation continued in M82 B until about 12--20 Myr ago.

The study of Melo et al. (2005) used {\sl HST}/Wide Field Planetary
Camera-2 observations to reveal 197 YMCs in the starburst core (with a
mean mass close to $2 \times 10^5 M_{\odot}$, largely independent of
the method used to derive these masses). Mayya et al. (2008) carried
out an objective search for star clusters on the {\sl HST}/ACS images
of M82 in the filters F435W ($B$), F555W ($V$), and F814W ($I$). Their
search led to the discovery of 393 clusters in the disk and 260
clusters in the nuclear region. Lim et al. (2013) used the same data
as the present study to analyze the age distribution and the
star-formation scenario in M82. Here, we focus on the (tidal)
disruption processes in the galaxy and how their influence has shaped
the surviving cluster population. This aspect has not yet been covered
in any detail for the galaxy as a whole, although a number of studies
have considered cluster disruption in spatially confined areas in M82
(e.g., de Grijs et al. 2003c for M82-B). In the present study, a
distance of 3.55 Mpc, or $(m - M)_0 = 27.75 \pm 0.07$ mag, is assumed,
as determined based on measurements of the $I$-band magnitude of the
tip of the red-giant branch (Lim et al. 2013). At the distance of M82,
$1''$ corresponds to a linear size of 17.2 pc.

\section{Source selection and photometry}
\label{reduction.sec}

\subsection{Original Data}

The main data set we used was obtained as part of the M82 Hubble
Heritage Program ({\sl HST} proposal GO-10776; PI: Mountain). In March
2006, the Hubble Heritage Team observed a large, four-color mosaic
image of M82 with the ACS/Wide Field Channel (WFC; pixel size $\sim
0.05''$) onboard the {\sl HST}, through the F435W, F555W, F814W, and
F658N filters. The latter is a narrow-band filter centered on the
H$\alpha$ emission line. The mosaic image is composed of six separate
subimages, with a negligible overlap area. These six `tiles' were
obtained with identical exposure times: each subimage is characterized
by four different exposure times of 1600, 1360, 1360, and 3320 s. All
subimages were combined to reduce the background noise and eliminate
cosmic rays (for details, see Mutchler et al. 2007)

A second data set, employing the filters F336W (equivalent to the
$U$-band filter; exposure times: 1050, 1215, and 1620 s), F110W
(roughly equivalent to a combination of the near-IR $Y$ and $J$ bands;
exposure times: 598 and 1195 s), and F160W (roughly similar to the
near-IR $H$ band; exposure times: 598 and 2395 s) has also been
included. These latter observations were obtained with the Wide Field
Camera 3 (WFC3; proposal GO-11360, PI: O'Connell). As opposed to the
images from the Hubble Heritage data set, which covers the entire
galaxy, these image sizes are smaller and only cover part of the
galaxy. For all filters, we constructed a combined set of images
covering $12288 \times 12288$ pixels$^2$, which are fully aligned and
have identical pixel sizes of $0.049''$; 1 pixel corresponds to
roughly 0.86 pc in linear size. Figure 1 of Lim et al. (2013) shows
the overlap areas of the Hubble Heritage data set and the additional
data sets composed of images in the F336W, F110W, and F160W filters.

\begin{table*}
 \begin{center}
 \begin{minipage}{180mm}
  \caption{Detailed observational properties of the adopted data sets.}
  \label{data.tab}
  \begin{tabular}{@{}lccccc@{}}
  \hline
   Filter   & Proposal ID/PI & Camera & Exposure time (s) &  Number of images & PHOTFLAM \\
 \hline
F435W ($B$) & GO-10776/Mountain  & {\sl HST}/ACS WFC & 1360 ($\times 2$), 1600, 3320 & 4 & $3.1412476\times10^{-19}$\\
F555W ($V$) & GO-10776/Mountain  & {\sl HST}/ACS WFC & 1360 ($\times 2$), 1600, 3320 & 4 & $1.9559270\times10^{-19}$\\
F814W ($I$) & GO-10776/Mountain  & {\sl HST}/ACS WFC & 1360 ($\times 2$), 1600, 3320 & 4 & $7.0723600\times10^{-20}$\\
F336W ($U$) & GO-11360/O'Connell & {\sl HST}/WFC3 & 1050, 1215, 1620 & 3 & $1.3407437\times10^{-18}$\\
F110W ($YJ$)& GO-11360/O'Connell & {\sl HST}/WFC3 & 598, 1195 & 2 & $1.5232975\times10^{-20}$\\
F160W ($H$) & GO-11360/O'Connell & {\sl HST}/WFC3 & 598, 2395  & 2 & $1.9106037\times10^{-20}$\\
\hline
\end{tabular}
\end{minipage}
\end{center}
\end{table*}

\subsection{Source Selection}

We used a custom-written procedure in IDL to find all relevant sources
for further analysis (cf. Barker et al. 2008). As input parameter, the
threshold value for detection is related to the number of sources
found in applying this procedure. We adopted our images in the {\sl
  HST}-equivalent $V$ and $I$ bands as our master images. To decide
which minimum (threshold) to use for source detection, for both master
images we determined the number of detections as a function of
detection threshold. We first estimated the sky (background) noise
level pertaining to the image frames in each filter (i.e., the
standard deviation in the background count level, determined in
regions of the image frames that were largely devoid of even extended
galaxy light), $\sigma_{\rm sky}$, using the {\sc imstat} routine of
the {\sc Iraf} package.\footnote{The Image Reduction and Analysis
  Facility ({\sc Iraf}) is distributed by the National Optical
  Astronomy Observatories, which is operated by the Association of
  Universities for Research in Astronomy, Inc., under cooperative
  agreement with the U.S. National Science Foundation. We used {\sc
    Iraf} version 2.15.1a (February 2011) for the data reduction
  performed in this study.} We next selected a range of thresholds
above the {\it local} background (sky + galaxy) level, expressed in
units of $\sigma_{\rm sky}$, i.e., [2, 3, 4, 5, 6]$\sigma_{\rm sky}$
to detect reliable sources. For lower thresholds, the resulting curves
are initially steep and then become shallower. We determined the
`knee' in the curve (see Barker et al. 2008), where real sources start
to dominate over noise, which occurs for a threshold value of
6$\sigma_{\rm sky}$ in both passbands. We find that if we vary the
local threshold by $\pm$1$\sigma_{\rm sky}$, the number of detected
sources varies by less than 5\%. These thresholds correspond to 0.20
and 0.8 counts s$^{-1}$ in the $V$ and $I$ bands, respectively,
resulting in catalogs containing 48,381 objects in the $V$ band and
129,091 objects in the $I$ band.

The next step involved cross-correlation of the source positions in
both filters to make sure that we are dealing with real objects in at
least these two filters. There are 22,467 sources in common to both
images; we adopted maximum offsets between source positions of one
pixel in each spatial dimension for a source pair to be considered a
match. We subsequently checked the source sizes. To define a minimum
size for (extended) cluster candidates, we generated artificial {\sl
  HST} point-spread functions (PSFs) using the {\sc TinyTim} package
(Krist \& Hook 1997) and, based on a PSF comparison, found that
sources with observed $\sigma_{\rm Gauss}\geq 0.84$ pixels are likely
to be genuine objects and not spurious features. This selection is
safe: the corresponding size cut of 0.84 pixels (intrinsic size
convolved with the PSF) is equivalent to Gaussian $\sigma$ of 0.7 pc
at the distance of M82. These (marginally to well-resolved) objects
must therefore represent genuine clusters instead of individual
stars. Inspection of the resulting size distribution shows that half
of the M82 cluster candidates are `point' sources, with $\sigma_{\rm
  Gauss}\leq 1.1$ pixels. The distribution shows an extended, flat
tail out to $\sigma_{\rm Gauss}\sim 5$ pixels.

\subsection{Photometry}

Using photometry tasks in IDL, we calibrated our source photometry by
applying the relevant zero-point offsets based on the values of the
image header keywords PHOTFLAM (see Table \ref{data.tab}). Using
3.5$\sigma_{\rm Gauss}$ for the source radii and [3.5--5]$\sigma_{\rm
  Gauss}$ for standard sky annulus radii (cf. de Grijs et al. 2013),
we obtained aperture photometry for 231 clusters out of 2320 extended
cluster candidates identified in total,\footnote{For comparison, Lim
  et al. (2013) found a total of 1105 star cluster candidates using a
  combination of automatic source detection and subsequent visual
  screening. Since that latter step involves a degree of subjectivity,
  we prefer to base our analysis entirely on automated methods, which
  we have used successfully in the past (e.g., de Grijs et
  al. 2003a,b, 2013; Barker et al. 2008; de Grijs \& Anders 2012),
  although we point out that a visual check done at an early stage of
  this work resulted in a cluster sample that led to statistically
  indistinguishable results compared with our automated methods. This
  is so, because our clusters are sufficiently massive so as not to be
  unduly affected by stochastic sampling at the fairly close distance
  to M82. We are striving to reduce the extent of subjectivity in our
  analysis methods by as much as possible.} which were present in all
of the $U, B, V, I$, and near-IR bands: see Table
\ref{nir.tab}. Compared with the $UBVI$ images, the near-IR data sets
are not very sensitive to faint sources, which hence limits the number
of objects detected in all filters. In order to have access to a large
cluster sample for our subsequent statistical analysis, we also
constructed a catalog containing the photometry of the 846 clusters
only found in the $U, B, V$, and $I$ bands but not at near-IR
wavelengths (see Table \ref{large.tab}); the limiting passband in this
case is the F336W filter. This latter sample constitutes our master
sample of extended star clusters in M82; the cluster properties are
included in Table \ref{large.tab}.

\begin{table*}
 \begin{center}
 \begin{minipage}{180mm}
  \caption{Photometry of the 231 M82 cluster candidates detected in
    all of the $UBVI(YJ)H$ passbands.}
  \label{nir.tab}
{\tiny
  \begin{tabular}{@{}cccccccccccccccc@{}}
  \hline
\multicolumn{2}{c}{R.A. (J2000)} & \multicolumn{2}{c}{Dec. (J2000)} & $m_{\rm F336W}$ & $m_{\rm F435W}$ & $m_{\rm F555W}$ & $m_{\rm F814W}$ & $m_{\rm F110W}$ & $m_{\rm F160W}$ \\
$(^{\circ})$ &(hh mm ss.ss)& $(^{\circ})$ & $(^{\circ} \, ' \, '')$ & (mag) & (mag) & (mag) & (mag) & (mag) & (mag) \\
 \hline
148.99374 & 09 55 58.50 & 69.685612 & 69 41 08.20 & $22.44 \pm 0.62$ & $20.82 \pm 0.17$ & $21.06 \pm 0.15$ & $21.45 \pm 0.12$ & $22.62 \pm 0.07$ & $22.65 \pm 0.07$ \\
148.99331 & 09 55 58.39 & 69.685324 & 69 41 07.17 & $22.17 \pm 0.56$ & $23.41 \pm 0.48$ & $23.27 \pm 0.37$ & $23.31 \pm 0.23$ & $22.51 \pm 0.16$ & $22.53 \pm 0.15$ \\
148.99044 & 09 55 57.71 & 69.684583 & 69 41 04.50 & $19.69 \pm 0.18$ & $20.51 \pm 0.16$ & $20.81 \pm 0.17$ & $22.07 \pm 0.15$ & $22.63 \pm 0.13$ & $22.66 \pm 0.13$ \\
148.98999 & 09 55 57.60 & 69.686996 & 69 41 13.19 & $24.29 \pm 1.49$ & $24.68 \pm 1.04$ & $24.21 \pm 0.56$ & $23.39 \pm 0.22$ & $26.11 \pm 0.44$ & $26.12 \pm 0.44$ \\
148.98958 & 09 55 57.50 & 69.684892 & 69 41 05.61 & $21.43 \pm 0.39$ & $22.25 \pm 0.42$ & $22.45 \pm 0.32$ & $23.38 \pm 0.31$ & $26.12 \pm 0.59$ & $26.12 \pm 0.59$ \\
$\cdots$  & $\cdots$    & $\cdots$  & $\cdots$    & $\cdots$         & $\cdots$         & $\cdots$         & $\cdots$         & $\cdots$         & $\cdots$         \\
\hline
\end{tabular}
}
\end{minipage}
\end{center}
\tablecomments{Table \ref{nir.tab} is published in its entirety in the
  electronic edition of {\it The Astrophysical Journal Supplements}. A
  portion is shown here for guidance regarding its form and content.}
\end{table*}

\begin{table*}
 \begin{center}
 \begin{minipage}{180mm}
  \caption{Observed and derived properties of the $UBVI$-based sample
    of candidate M82 clusters.}
  \label{large.tab}
{\tiny
  \begin{tabular}{@{}cccccccccccccccc@{}}
  \hline
\multicolumn{2}{c}{R.A. (J2000)} & \multicolumn{2}{c}{Dec. (J2000)} & $m_{\rm F336W}$ & $m_{\rm F435W}$ & $m_{\rm F555W}$ & $m_{\rm F814W}$ & $\log(t)$ & $\log(M_{\rm cl})$ & $E(B-V)$ & $R_{\rm hl}$ \\
$(^{\circ})$ &(hh mm ss.ss)& $(^{\circ})$ & $(^{\circ} \, ' \, '')$ & (mag) & (mag) & (mag) & (mag) & [yr] & [$M_\odot$] & (mag) & (pc) \\
 \hline
148.87847 & 09 55 30.83 & 69.665917 & 69 39 57.30 & $24.22 \pm 1.41$ & $24.18 \pm 0.64$ & $23.38 \pm 0.33$ & $23.47 \pm 0.20$ & $8.55_{-0.33}^{+0.47}$ & $5.58_{-0.36}^{+0.45}$ & 0.00 & 2.52 \\
148.88296 & 09 55 31.91 & 69.670109 & 69 40 12.39 & $28.57 \pm 1.88$ & $25.77 \pm 1.36$ & $25.14 \pm 0.75$ & $24.20 \pm 0.27$ & $8.49_{-0.46}^{+0.68}$ & $5.05_{-0.52}^{+0.66}$ & 1.00 & 0.83 \\
148.89413 & 09 55 34.59 & 69.676876 & 69 40 36.75 & $24.25 \pm 1.43$ & $23.68 \pm 0.50$ & $23.77 \pm 0.39$ & $24.29 \pm 0.27$ & $7.76_{-0.52}^{+0.58}$ & $4.70_{-0.56}^{+0.62}$ & 0.30 & 2.87 \\
148.89483 & 09 55 34.76 & 69.671262 & 69 40 16.54 & $29.11 \pm 1.88$ & $26.35 \pm 1.86$ & $25.65 \pm 0.99$ & $24.13 \pm 0.27$ & $8.21_{-0.62}^{+0.65}$ & $5.13_{-0.71}^{+0.73}$ & 1.30 & 0.97 \\
148.89570 & 09 55 34.97 & 69.670507 & 69 40 13.83 & $24.29 \pm 1.46$ & $23.63 \pm 0.52$ & $24.11 \pm 0.49$ & $24.35 \pm 0.34$ & $7.72_{-0.29}^{+0.34}$ & $3.96_{-0.34}^{+0.35}$ & 0.30 & 3.41 \\
 $\cdots$  & $\cdots$    & $\cdots$  & $\cdots$    & $\cdots$           & $\cdots$           & $\cdots$           &$\cdots$    &$\cdots$&$\cdots$&$\cdots$&$\cdots$\\
\hline
\end{tabular}
}
\end{minipage}
\end{center}
\tablecomments{Typical uncertainties in the total (foreground +
  internal) extinction, $E(B-V)$ are 0.05 to 0.10 mag; typical
  uncertainties in $R_{\rm hl}$ are 0.02 to 0.05 pc.\\ Table
  \ref{large.tab} is published in its entirety in the electronic
  edition of {\it The Astrophysical Journal Supplements}. A portion is
  shown here for guidance regarding its form and content.}
\end{table*}

\subsection{Completeness Analysis}

In order to obtain a reliable number estimate, we need to assess how
many objects may have been missed by our processing technique. We
first generated a blank template image without any background noise,
with the same size as that of the Hubble Heritage data set. Then, we
used the {\sc mkobj} task in {\sc Iraf} to generate 500 artificial
sources with $\sigma_{\rm Gauss} = 1.0$ pixels (representing `point'
sources), which we added to $500 \times 500$ pixel$^2$ sections of the
blank template at random $(X, Y)$ coordinates, ranging from 0 to 500
pixels in both dimensions. All 500 artificial sources were assigned
the same magnitude; we repeated this approach by varying the
magnitudes of the artificial sources from 18.0 mag to 28.0 mag for
each filter, in steps of 0.5 mag. We added all artificial source
frames to both the template and science images. We repeated this
procedure to cover the entire template region, thus allowing us to
gain a sense of the variability of the completeness levels across the
full observational area.

\begin{figure}[ht!]
\begin{center}
\includegraphics[width=1.1\columnwidth]{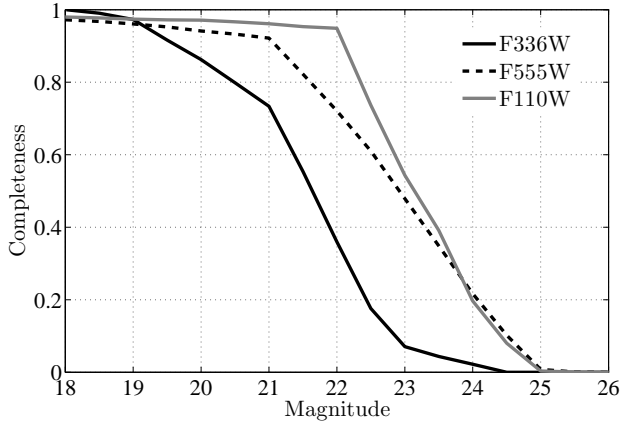}
\caption{Completeness curves for the cluster sample obtained from the
  {\sl HST}/WFC3 and ACS images. The analysis was done for all data in
  the F336W, F555W, and F110W filters.}
\label{F2}
\end{center}
\end{figure}

Finally, we `discovered' the number of artificial sources in both the
template and science images, using the same approach for both. (We
similarly applied our source discovery routine to the science images
to find the real clusters in Section 2.2.) Since the template image
contained no background noise, we can recover most input artificial
sources, although some artificial sources may have disappeared into
the physical noise of the science images, depending on the
combinations of their integrated magnitudes and extent, as well as on
the real background variations. In addition, saturated sources and
blending may cause failures to recover the artificial sources. The
template image hence became a perfect comparison sample to check how
many sources should be recovered under ideal conditions. We counted
the number of recovered artificial sources (from the science image)
and the number of recovered artificial sources (from the template) to
estimate the completeness, i.e. $f_{\rm comp} = N_{\rm rec}/N_{\rm
  tot}$, where $N_{\rm rec}$ is the number of artificial objects we
recovered and $N_{\rm tot}$ is the number of input objects; $f_{\rm
  comp}$ was simultaneously corrected for the effects of blending and
saturation, as well as background noise. We show the results for the
full science observations, as a function of magnitude, in Figure
\ref{F2}. We found that, for the observational data set as a whole, at
approximately 21.5 mag in the F336W band, 23 mag in F555W, and 23.5
mag in the F110W band, the completeness fraction drops to below
50\%. This indicates that at those magnitudes, the background noise
starts to dominate the image quality. In the remainder of this paper,
we will only consider objects above their respective 50\% completeness
limits and correct the relevant numbers for sampling
incompleteness. Note that, strictly speaking, this relates to the
cluster subsample representing point sources, which thus implies that
the completeness levels derived here are upper limits to the
complement of more extended cluster candidates. However, in practice
the completeness levels derived here differ only marginally (and
firmly within the photometric uncertainties) from those for objects
with $\sigma_{\rm Gauss} \in [2,3]$ pixels; only $\sim 15$\% of M82
cluster candidates are more extended.

\section{Cluster evolution}
\subsection{Parameter Determination}

In Section \ref{reduction.sec} we obtained the observed (broad-band)
spectral-energy distributions (SEDs). A small fraction of our sample
(231 clusters) were covered in all of the $U, B, V, I$, and near-IR
bands, whereas a larger cluster sample of 846 objects was found only
in the $UBVI$ bands. We used the {\sc galev} evolutionary synthesis
models (Kotulla et al. 2009, and references therein;
http://www.galev.org) to determine the best-fitting values for the
ages, masses, extinction values, and metallicities for all clusters by
comparing the set of model SEDs with our observed SEDs. Our
statistical comparison was based on application of the {\sc AnalySED}
tool, which has been tested extensively both internally (de Grijs et
al. 2003a,b; Anders et al. 2004b) and externally (de Grijs et
al. 2005), using both theoretical and observed young to
intermediate-age ($\leq 3 \times10^9$ yr-old) star cluster SEDs.

\begin{figure}[ht!]
\begin{center}
\includegraphics[width=1.1\columnwidth]{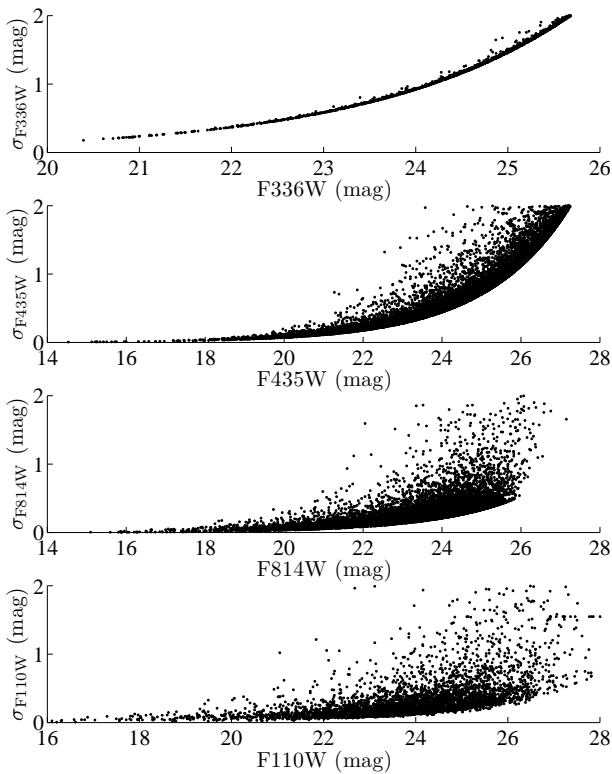}
\caption{Photometric uncertainties as a function of integrated cluster
  magnitude for the F336W, F435W, F814W, and F110W filters.}
\label{F7}
\end{center}
\end{figure}

The relative ages and masses within a given cluster system can be
determined to very high accuracy, provided that a combination of at
least four passbands is used as input parameters (Anders et
al. 2004b). We first applied the {\sc AnalySED} approach to our
smaller set of broad-band $UBVI(YJ)H$ cluster SEDs to determine the
best-fitting overall metallicity, which yielded $Z=0.02$ (solar
metallicity). Based on abundance measurements of H{\sc ii} regions and
young clusters in M82 (Smith et al. 2006; Lan\c{c}on et al. 2008),
solar metallicity for the star clusters seems indeed reasonable. We
thus decided to fix the model cluster metallicities to the solar
value, leaving both the cluster ages and extinction values as free
parameters. The advantage of this approach is that this leaves us with
one fewer free parameter to determine, which in turn renders our
resulting age, mass, and extinction estimates more robust. The shape
of the broad-band SEDs constrains the ages and extinction values,
whereas the absolute flux levels result in the corresponding cluster
masses. Despite the sometimes significant photometric uncertainties
(see Fig. \ref{F7}), generally caused by the highly variable
background, our SED-matching approach successfully converged to a
reasonably well-determined set of ages (cf. Table \ref{large.tab}).

\begin{figure}[ht!]
\begin{center}
\includegraphics[width=1.1\columnwidth]{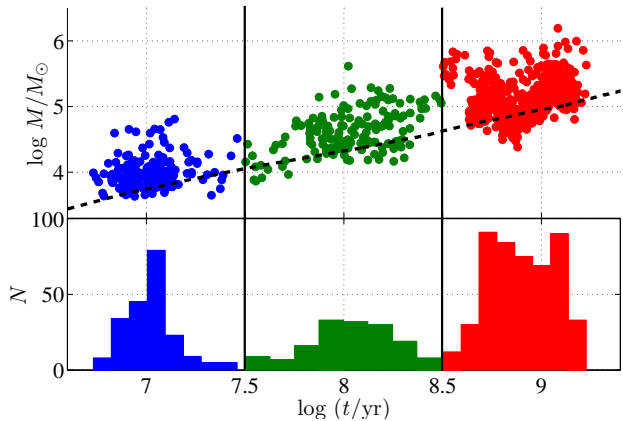}
\caption{Top: Distribution of M82 clusters in the log(age)--log(mass)
  plane. The dashed line indicates the F336W band's 50\% completeness
  level (for `point' sources). Blue, green, and red data points
  indicate the youngest, intermediate-age, and oldest cluster samples,
  respectively. Bottom: Corresponding age distribution.}
\label{F3}
\end{center}
\end{figure}

The F336W band is the filter limiting our completeness analysis; it
reaches the 50\% detection limit at a magnitude of 21.5. Although the
M82 starburst regions are pervaded by high-extinction filaments, the
outermost parts of the starburst core have lower extinction and can be
studied with optical telescopes. However, the use of optical
wavelengths, and in particular the need for a sufficiently high
signal-to-noise ratio in the F336W filter, effectively limits our
sampling of the M82 starbursts to their surface regions. Using the
{\sc galev} model suite, we draw the 50\% completeness limit (for
`point' sources) defined by the F336W band in the log(age)--log(mass)
diagram of Fig. \ref{F3} (top). Formally, 691 of our 846 clusters are
located above the F336W 50\% completeness limit. In essence, our
cluster sample therefore represents a `$U$-band selected' sample. With
this caveat in mind, two clear ``gaps'' in the clusters' galaxy-wide
age distribution become apparent, which implies a bursty
cluster-formation history across the face of the galaxy. The
corresponding histogram (see Fig. \ref{F3}, bottom) shows that the gap
at the youngest age is located at ${\log}(t \mbox{ yr}^{-1}) \sim
7.5$, while the second gap is located at approximately ${\log}(t
\mbox{ yr}^{-1}) = 8.5$.

We hence divided our star clusters into three samples: the young
sample is composed of clusters with ages younger than $10^{7.5}$ yr
($t \le 30$ Myr), the intermediate-age sample's member clusters are
aged between $10^{7.5}$ yr and $10^{8.5}$ yr ($30 \mbox{ Myr} \le t
\le 300$ Myr), and the old sample hence consists of the remaining
clusters, characterized by ages in excess of $10^{8.5}$ yr ($t \ge
300$ Myr). The youngest sample thus encompasses the time of the most
recent starburst in the galaxy's nucleus, 4--6 Myr ago (e.g., Barker
et al. 2008), while the intermediate-age sample may be associated with
the burst of cluster formation induced by the last M81/M82 flyby
(e.g., de Grijs et al. 2001; Smith et al. 2007).

Lim et al. (2013) separated their sample of more than 1100 clusters by
region, and found that each region (specifically, the galaxy's disk,
its halo, the nuclear region, and region M82-B) exhibits a different
cluster formation history. Here, we are interested in the global
characteristics of the M82 cluster population. Somewhat surprisingly
given the good match between their and our sets of age derivations
(see below), Lim et al.'s (2013) overall cluster age distribution
exhibits a clear, dominant peak at $\log( t \mbox{ yr}^{-1}) \sim
8.7$, which is consistent with the age distribution shown in
Fig. \ref{F3} (bottom); except for the nuclear region, they do not
show evidence of any gaps in the age distribution. In contrast, de
Grijs et al. (2003c; their Fig. 4c) found a significant gap in the
cluster-formation rate in M82-B near ${\log}(t \mbox{ yr}^{-1}) \sim
8.5$, a region not selected for specific analysis by Lim et
al. (2013).

We compared our photometry and derived cluster ages and masses with
the equivalent values of Melo et al. (2005), Smith et al. (2007), and
Lim et al. (2013) for 21, eight, and 311 clusters in common,
respectively. A quantitative comparison between our values and the
relevant samples from the literature is given in Table
\ref{comparison.tab}.

The majority of the clusters in the intermediate-age and old samples
are more massive than a few $\times 10^4 M_\odot$. Based on our recent
analysis of the impact of stochastic sampling of stellar mass
functions on integrated star cluster properties (Anders et al. 2013;
de Grijs et al. 2013; see also Silva-Villa \& Larsen 2011) we conclude
that the effects of stochasticity are minimal for M82 clusters that
are older than a few $\times 10^7$ yr. The context for the clusters in
our youngest subsample is different, however: these clusters are all
less massive than $10^5 M_\odot$. Nevertheless, the young ages of a
subset of these clusters have been confirmed independently on the
basis of spectroscopic observations ({\it nuclear region}: Smith et
al. 2006; Westmoquette et al. 2007; {\it M82-B}: Konstantopoulos et
al. 2008), as well as by independent analysis of the cluster
population in the starburst nucleus (Barker et al. 2008), which also
included information about the clusters' H$\alpha$ luminosities. In
addition, the match of our cluster age estimates with the largest
comparison sample, that of Lim et al. (2013), is equally good for any
age range. This supports our assumption that stochasticity in the
clusters' mass functions does not {\it dominate} the results presented
here (see also the equivalent conclusion in de Grijs et al. 2013),
although some level of stochasticity is likely to affect both our and
Lim et al.'s (2013) results.

\begin{table}
 \begin{center}
  \caption{Statistical comparison of our photometry and derived
    parameters with respect to a number of key comparison samples. All
    differences and standard deviations are in the sense ``our
    measurements minus literature data''; photometric comparisons are
    in magnitudes.}
    \label{comparison.tab}
  \begin{tabular}{@{}lccc@{}}
  \hline
Parameter & \multicolumn{3}{c}{Differences with respect to}\\
\cline{2-4}
          & Melo          & Smith         & Lim \\
          & et al. (2005) & et al. (2007) & et al. (2013) \\
\hline
$N_{\rm cl}$                      & 21       & 8     & 311    \\
\hline
\multicolumn{2}{l}{\bf 1. Photometry} \\
$\Delta m_{\rm F336W}$            &                  &       & $\;\;\;$0.0008 \\
$\sigma_{\rm F336W}$              &                  &       & $\;\;\;$0.0348 \\
$\Delta m_{\rm F435W}$            & $-0.010$         & 0.025 & $\;\;\;$0.0007 \\
$\sigma_{\rm F435W}$              & $\;\;\;$0.245    & 0.212 & $\;\;\;$0.0373 \\
$\Delta m_{\rm F555W}$            & $-0.010$         & 0.088 & $\;\;\;$0.0005 \\
$\sigma_{\rm F555W}$              & $\;\;\;$0.175    & 0.203 & $\;\;\;$0.0422 \\
$\Delta m_{\rm F814W}$            & $-0.095$         & 0.100 & $\;\;\;$0.0002 \\
$\sigma_{\rm F814W}$              & $\;\;\;$0.235    & 0.212 & $\;\;\;$0.0518 \\
$\Delta m_{\rm F110W}$            &                  &       & $-0.0001$ \\
$\sigma_{\rm F110W}$              &                  &       & $\;\;\;$0.0501 \\
\hline
\multicolumn{2}{l}{\bf 2. Derived parameters} \\
$\Delta \log(M_{\rm cl}/M_\odot)$ & 0.085    &       &        \\
$\sigma_{\log M_{\rm cl}}$        & 0.091    &       &        \\
$\Delta \log(t \mbox{ yr}^{-1})$  &          & 0.052 & 0.1668 \\
$\sigma_{\log t}$                 &          & 0.074 & 0.2135 \\
\hline
\end{tabular}
\end{center}
\end{table}

\begin{figure}[ht!]
\begin{center}
\includegraphics[width=1.1\columnwidth]{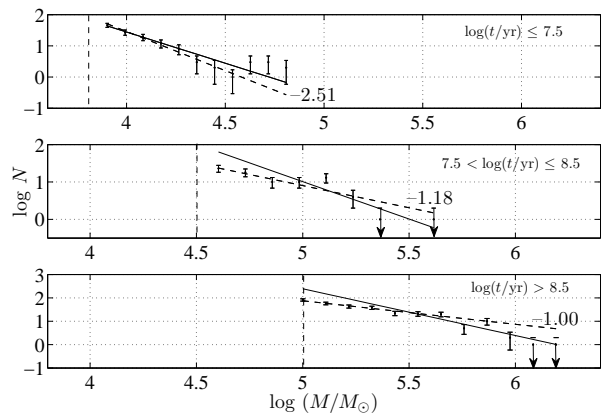}
\caption{CMFs for the three different age ranges adopted in this
  study. The dashed lines show the best-fitting power-law CMFs to the
  full mass ranges available. The solid lines indicate the
  theoretically predicted power-law CMF. The vertical dashed lines
  reflect the samples' 50\% completeness limits.}
\label{F4}
\end{center}
\end{figure}

Figure \ref{F4} shows the mass distribution of the clusters located
above the 50\% (F336W) completeness limits (for `point' sources) for
all three subsamples. In each panel, the CMF can be reasonably well
fitted by a power-law CMF of the form $\log N_{\rm cl} \propto
\log(M_{\rm cl})^{-\alpha}$ (dashed lines). The slopes of the dashed
lines (obtained by fitting all clusters in the respective samples) are
$\alpha = (-2.51 \pm 0.41), (-1.18 \pm 0.43)$, and $(-1.00 \pm 0.18)$
for the young, intermediate-age, and old samples, respectively. The
solid lines indicate the theoretically expected slope of $\alpha = -2$
(e.g., Whitmore et al. 1999; Zhang \& Fall 1999).

Our results in Fig. \ref{F4} display gradually flatter CMFs
(pertaining to the full, observed mass ranges) with increasing cluster
age. To distinguish between the ``low-mass'' and ``high-mass'' ranges,
we divide all three samples at the mass where their CMFs exhibit a
clear break in their slopes, which occurs at ${\log} (M_{\rm
  cl}/M_{\odot}) \sim (4.6 \pm 0.1), (5.2 \pm 0.2)$, and $(5.7 \pm
0.2)$ for the young, intermediate-age, and old samples,
respectively. We will now first discuss the intermediate-age---${\log}
(t \mbox{ yr}^{-1}) \in [7.5, 8.5]$---and old samples, ${\log} (t
\mbox{ yr}^{-1}) \geq 8.5$. If we only consider the less massive
clusters---i.e., for the intermediate-age sample, we select the four
lowest-mass bins, ${\log} (M_{\rm cl}/M_{\odot}) \le 5$, and for the
old sample we select the first seven bins, ${\log} (M_{\rm
  cl}/M_{\odot}) \leq 5.61$---their distributions are (perhaps
surprisingly given the expected effects of stochasticity) relatively
similar, with slopes of $\alpha = (-1.18 \pm 0.43)$ and $\alpha =
(-1.00 \pm 0.18)$ for the intermediate-age and old samples,
respectively. These slopes are significantly shallower than the
equivalent slope pertaining to the young sample---${\log} (t \mbox{
  yr}^{-1}) \leq 7.5$---of $\alpha= -2.51 \pm 0.41$ for clusters with
${\log} (M_{\rm cl}/M_{\odot}) \leq 4.54$.

Despite their different mass ranges, the intermediate-age and old
samples exhibit a high similarity, i.e., relatively shallow CMFs over
a small, low-mass range (presumably caused by the effects of cluster
disruption; see below), combined with a steeper slope pertaining to
the least-evolved, most massive clusters. For the intermediate-age
sample, the fit to the CMF using the four highest-mass bins, where the
effects of stochastic sampling are expected to be negligible
(cf. Anders et al. 2013)---${\log} (M_{\rm cl}/M_{\odot}) \geq
5.0$---yields a relatively poorly constrained slope of $\alpha \simeq
-2.17$. Meanwhile, the fit to the old sample's CMF defined by the five
highest-mass bins---${\log} (M_{\rm cl}/M_{\odot}) \geq
5.6$---indicates a slope of $\alpha = -2.21 \pm 0.42$. However, if we
consider the young sample and only select its three highest-mass
bins---${\log}(M_{\rm cl}/M_{\odot}) \geq 4.5$---the resulting CMF is
quite shallow, $\alpha \sim -1.0$. We note that this is likely the
result of small-number statistics, given that the young sample
contains only a handful of clusters with masses ${\log}(M_{\rm
  cl}/M_{\odot}) \geq 4.5$. Nevertheless, the masses of the most
massive clusters in the young sample define an upper envelope in
Fig. 3 that extends from the youngest to the oldest ages, following a
trend that is usually attributed to size-of-sample effects (e.g.,
Hunter et al. 2003). The lower-mass range of the young clusters' CMF
exhibits a much steeper CMF that is roughly consistent with the
theoretically predicted initial, unevolved CMF, characterized by
$\alpha = -2$. In this context, both the intermediate-age and old
samples display shallow-to-steep CMF slope changes when going from
low- to high-mass clusters, which is most likely owing to the
preferential disruption of the lower-mass clusters (see below).

This distinct difference with respect to the less massive clusters'
CMF in the young cluster sample may indicate a mass-specific
disruption law: clusters that are less massive than a critical mass
are preferentially disrupted after a typical timescale on which
dynamical disruption sets in. Since the cluster-formation history
across the galaxy has been bursty, we cannot simply apply ``standard''
cluster disruption analysis (e.g., Boutloukos \& Lamers 2003; de Grijs
et al. 2003c) to derive this timescale.

\subsection{Tidal-disruption processes}

While the young cluster sample exhibits an excess of relatively
massive clusters (which may be owing to small-number statistics)---and
hence displays a flatter distribution for the highest-mass range,
i.e., for $\log(M_{\rm cl}/M_\odot) \ga 4.5$---both the
intermediate-age and old samples exhibit apparently steeper slopes
consistent with $\alpha = -2$ at the high-mass end, i.e., for
$\log(M_{\rm cl}/M_\odot) \ga 5.1$ and $\log(M_{\rm cl}/M_\odot) \ga
5.7$, respectively, than those defined by the lower-mass clusters in
the respective samples. This implies that the massive star clusters in
these subsamples are less significantly affected by disruption
processes than their lower-mass counterparts, as expected from the
mass-dependent cluster disruption formalism originally developed by
Boutloukos \& Lamers (2003) and expanded upon since by a number of
authors (e.g., de Grijs et al. 2003a,c; Lamers et al. 2005a,b; Gieles
et al. 2005, 2007; Parmentier \& de Grijs 2008).

\begin{figure}[ht!]
\begin{center}
\includegraphics[width=\columnwidth]{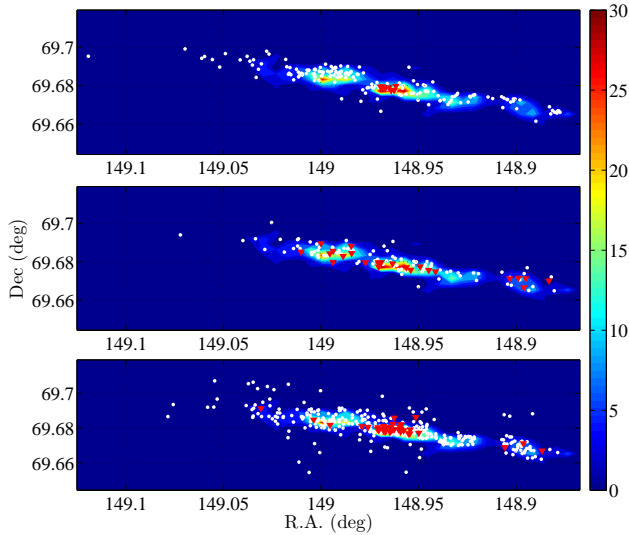}
\caption{(top) Spatial distribution of the young cluster sample (white
  points). The most massive young clusters---${\log} (M_{\rm
    cl}/M_{\odot}) \geq 4.54$---are indicated by red triangles; the
  background colors represent the number density distribution of all
  detected sources. (middle and bottom) Same as the top panel, but for
  the intermediate-age and old samples, respectively.}
\label{F5}
\end{center}
\end{figure}

We first investigate if the spatial distribution of the most massive
clusters in the young sample is different from those of the
intermediate-age and old samples. In Fig. \ref{F5} (top) we show the
spatial distribution of all young star clusters (white points), as
well as that of the young star clusters with ${\log} (M_{\rm
  cl}/M_{\odot}) \geq 4.5$ (red triangles). In order to compare these
to the total sample (including all detected sources), we have also
included the number densities, indicated by the background colors. We
find that compared with the total young cluster sample, the most
massive young clusters are all located within a very compact
region. In the middle and bottom panels, we display the same spatial
distributions for the intermediate-age and old samples,
respectively. Quantitatively, the most massive young clusters are
found in a region spanning $8.3'' \times 28''$ in right ascension and
declination, while their intermediate-age counterparts cover an
equivalent area of approximately $460'' \times 83''$. The distribution
of the most massive clusters---${\log} (M_{\rm cl}/M_{\odot}) \geq
4.99$, red triangles---in the intermediate-age sample compared with
that of their counterparts in the old sample, is significantly
different; the former are dispersed around the whole galaxy. The 30
most massive clusters in the old sample are more centrally
concentrated, covering a range of approximately $105'' \times 20''$
(right ascension $\times$ declination); the remaining eight clusters
are found throughout the galaxy's disk.

The old sample represents the potential field of the oldest galactic
components. This sample contributes most clusters among all detected
sources (55\%; 378 of the total of 691 clusters above the prevailing
completeness threshold adopted), and hence their distribution is
strongly associated with the high background number-density region. We
propose the following evolutionary scenario. The oldest clusters
likely formed at the time when the galaxy itself was in the process of
formation, so the gravitational background experienced by this first
cluster population at the time of their formation must have been very
different from that in which subsequent cluster populations
formed. When these latter populations formed, they found themselves
inside a well-developed, centrally peaked gravitational potential,
which would have strongly affected the evolution of these subsequently
formed cluster populations. In the number-density color figure, the
highest-density region was likely the preferential nursery of the
descendants of both the intermediate-age and old clusters (i.e., a
nursery of both the young and intermediate-age star clusters,
respectively), but in the mean time, the strong tidal field in the
highest-density region will also have tidally truncated and stripped
off the outer layers of the largest clusters, thus only leaving the
more compact clusters as survivors.

In order to test if this proposed scenario is viable, we explored the
size distribution of the surviving, most massive young clusters. If
those young massive star clusters really have been affected by tidal
disruption, they should be significantly smaller than their
counterparts that are located far beyond the immediate influence of
the central tidal field. We used the {\sc ishape} software package
(Larsen 1999) to determine the cluster sizes. This package is used to
estimate the intrinsic shape parameters of extended objects given the
observational point-spread function. It attempts to fit the radial
profiles of extended sources with simple analytical functions. Here,
we are mainly concerned with the observed clusters' empirical
sizes. The derived effective radii are not very sensitive to the
choice of model (Larsen 1999). A related result was obtained by Kundu
\& Whitmore (1998), who found that effective radii derived from fits
to a King (1962) model were also quite insensitive to the adopted
concentration parameter $c\equiv{\rm log}(r_{\rm t}/r_{0})$, where
$r_{\rm t}$ is the tidal radius and $r_{0}$ the King radius. We
adopted a King model with $c = 5$ to fit the clusters' radial
profiles. Finally, we determined the clusters' half-light radii
($R_{\rm hl}$) as an appropriate representation of their sizes, in
units of pixels.\footnote{Size comparisons for clusters of different
  ages may introduce a bias because of mass segregation (de Grijs et
  al. 2002a,b,c), however, which will cause $R_{\rm hl}$ to decrease
  as a star cluster evolves. At the same time, their half-{\it mass}
  radii will increase, so that for relatively old sample clusters, we
  may overestimate their average densities.}

\begin{figure}[ht!]
\begin{center}
\includegraphics[width=\columnwidth]{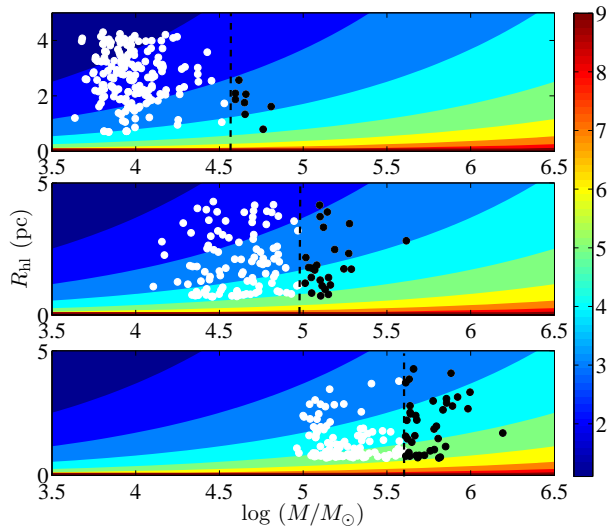}
\caption{(top) ${\log} (M_{\rm cl}/M_{\odot})$ versus $R_{\rm hl}$
  diagram for the young star clusters. The vertical dashed line
  indicates the critical mass adopted for the most massive clusters,
  i.e., ${\log} (M_{\rm cl}/M_{\odot}) \geq 4.54$, while the
  background colors and contours indicate the cluster population's
  density distribution. (middle and bottom) Same as the top panel, but
  for our intermediate-age and oldest samples, respectively.}
\label{F6}
\end{center}
\end{figure}

Figure \ref{F6} (top) displays the ${\log} (M_{\rm cl}/M_{\odot})$
versus $R_{\rm hl}$ diagram for all young star clusters. The dashed
line indicates the critical mass we adopted for the most massive
clusters in the young sample. An apparent mean smaller size for these
most massive clusters, $\langle R_{\rm hl} \rangle \simeq 2.6$ pc, is
found, as well as a smaller size spread: the less massive clusters
have a mean half-light radius of $\langle R_{\rm hl} \rangle \simeq
4.3$ pc. We also calculated the clusters' average density, $\rho =
M_{\rm cl}/(2R_{\rm hl}^{3})$ (see the background colors), where we
simply assumed that the clusters' half-light radii are roughly equal
to their half-mass radii ($R_{\rm hm}$). It thus appears that the most
massive young clusters are also the highest-density, most compact
objects. Note that despite the relatively small sample size of eight
objects above the critical mass limit, these eight sources are the
brightest clusters in the young subsample. In Fig. \ref{F3}, they are
found well above our galaxy-wide selection limit. If the young sample
had included additional, more extended objects of similar masses, our
procedures would, without a doubt, have detected such objects. This is
supported by the fact that the young sample {\it does} include more
extended objects below the critical mass limit, which are less
luminous (by definition, given their very similar ages) than the
higher-mass objects we would have missed. As such, we conclude that
there is indeed a tantalizing hint that the most massive young
clusters are systematically smaller than their lower-mass
counterparts. (Of course, this argument hinges on the basic underlying
assumption that the clusters' radial profiles do not differ
significantly among the sample objects.)

In Fig. \ref{F6} (middle and bottom), we show the same distributions
as in Fig. \ref{F6} (top), but for the intermediate-age and old
samples. This time, however, we do not observe any clearly smaller
mean size for the most massive clusters: for the intermediate-age
clusters, the mean half-light radii for the low- and high-mass
subsamples are $\langle R_{\rm hl,low} \rangle \simeq 4.3$ pc and
$\langle R_{\rm hl,high} \rangle \simeq 4.2$ pc, respectively. The
equivalent sizes for the the old sample are 4.1 pc and 4.3 pc,
respectively. This may indicate that the remaining massive star
clusters in the intermediate-age and old samples are not as strongly
affected by the prevailing tidal field, in part because of their more
extended distributions, i.e., away from the peak of the tidal field's
distribution. This is a reasonable conclusion, since the first
generation of star clusters (the old sample) likely formed when the
galaxy itself was still forming, so that (i) a strong tidal field may
not yet have been established, and (ii) the remaining old clusters
have proved robust with respect to the effects of the prevailing tidal
field. We are thus likely observing an old cluster population that has
survived the potentially devastating effects of tidal stripping. Note,
however, that although the intermediate-age, most massive clusters are
more dispersed than their counterparts in the young sample, which
should not yet have been affected strongly by the tidal field, some
may already have been disrupted.

Alternatively, the reason why the massive old clusters may still be
confined to a relatively small volume in and near the galactic center
region might be found in the complex structure of the M82 disk. It is
well-known that the inner $\sim 1$ kpc of M82 is dominated by a
stellar bar (e.g., Wills et al. 2000) in solid-body rotation. The
phase mixing might be slow enough for the bar to retain its identity
over a sufficient time so as to remain self-constrained
(J. S. Gallagher, private communication). In addition, since the
density in the region is high, simple calculations imply that the
area's self-gravity is non-negligible compared with the rotational
shear, therefore also prohibiting a rapid dispersion of the central
starburst region.

Finally, we note that the old clusters survived the recent tidal
encounter with M81, which may have occurred between approximately 150
Myr and 1 Gyr ago (e.g., Brouillet et al. 1991; Yun et al. 1994; de
Grijs et al. 2001, 2003c; Smith et al. 2007). Although this might
appear curious at first sight, we point out that our old cluster
sample only includes objects that are more massive than a few $\times
10^4 M_\odot$, while disruption -- both owing to internal relaxation
and caused by external perturbations -- predominantly affects the
lower-mass clusters in a given sample. However, because of the
age-dependent selection limit, we cannot make any definite statements
about whether or not the last tidal encounter with M82 may have led to
significantly enhanced cluster disruption among the lower-mass
clusters in the oldest subsample (but, for a similar scenario in the
context of the interacting galaxy M51, see Lamers et al. 2005a).

\section{Conclusion}

Using high-resolution {\sl HST} images based on multiple-passband
observations, we selected a complete sample of 846 star clusters in
the disk of M82, for which we determined the respective magnitudes, as
well as their age, mass, and size information. We aimed at
characterizing the effects of cluster disruption based on a detailed
analysis of the clusters' age and mass distributions. Our primary
results can be summarized as follows.

\begin{enumerate}
\item The young sample---${\log} (t \mbox{ yr}^{-1}) \leq
  7.5$---displays a steep-to-shallow CMF slope change when going from
  low to high cluster masses, with a characteristic break near ${\log}
  (M_{\rm cl}/M_\odot) = 4.6$, while both the
  intermediate-age---${\log} (t \mbox{ yr}^{-1}) \in [7.5, 8.5]$---and
  old samples---${\log} (t \mbox{ yr}^{-1}) \geq 8.5$---display a
  shallow-to-steep slope change in the same direction. This distinct
  difference with respect to the less massive clusters' CMF in the
  young cluster sample may indicate a mass-specific disruption law:
  clusters that are less massive than a critical mass are
  preferentially disrupted on a typical, mass-dependent timescale
  ($t_{\rm dis} \gg 30$ Myr).

\item Compared with the total young cluster sample, the most massive
  young clusters are all located within a very compact spatial
  region. The distribution of the most massive clusters in the
  intermediate-age sample---${\log} (M_{\rm cl}/M_{\odot}) \geq
  4.99$---compared with that of their counterparts in the young
  sample, is markedly different; the latter are dispersed around the
  whole galaxy. The 30 most massive clusters in the old sample are
  significantly more centrally concentrated.

\item It is clear that the most massive young clusters are also the
  highest-density, most compact objects. This result hence strongly
  suggests that the properties of the most massive young clusters are
  affected by the galaxy's tidal field. For the intermediate-age and
  old samples, we do not observe any clearly smaller size for the most
  massive clusters. This may indicate that we are observing a
  different subset of the initial cluster population for each of the
  age ranges (i.e., differently affected by tidal effects and
  evolution), and that the remaining, surviving massive star clusters
  in the intermediate-age and old samples are---on the whole---more
  robust with respect to changes due to the prevailing tidal field.

\end{enumerate}

\section*{Acknowledgments}

This article forms part of the first author's requirements to obtain
an M.Sc. degree at Peking University. We thank Chaojian Wu for
technical assistance with programming issues. We acknowledge research
support from the National Natural Science Foundation of China (grants
11073001 and 11373010).

\end{document}